\documentclass[aps,pra,twocolumn,superscriptaddress,showpacs]{revtex4-1}
\usepackage[utf8]{inputenc}
\usepackage{amsmath}
\usepackage{amsfonts}
\usepackage{amssymb}
\usepackage{graphicx}
\usepackage{dcolumn}

\begin{document}

\title{Cross-damping effects in 1S-3S spectroscopy\\ of hydrogen and deuterium}

\author{Hélène Fleurbaey}
\author{François Biraben}
\author{Lucile Julien}
\affiliation{Laboratoire Kastler Brossel, UPMC-Sorbonne Universités, CNRS, 
ENS-PSL Research University, Collège de France, 4 place Jussieu, Case 74,
75252 Paris Cedex 05, France}
\author{Jean-Philippe Karr}
\affiliation{Laboratoire Kastler Brossel, UPMC-Sorbonne Universités, CNRS, 
ENS-PSL Research University, Collège de France, 4 place Jussieu, Case 74,
75252 Paris Cedex 05, France}
\affiliation{Université d’Evry-Val d’Essonne, Boulevard François Mitterrand, 
91000 Evry, France}
\author{François Nez}
\affiliation{Laboratoire Kastler Brossel, UPMC-Sorbonne Universités, CNRS, 
ENS-PSL Research University, Collège de France, 4 place Jussieu, Case 74,
75252 Paris Cedex 05, France}

\date{\today}

\begin{abstract}
We calculate the cross-damping frequency shift of a laser-induced two-photon transition 
monitored through decay fluorescence, by adapting the analogy with Raman scattering 
developed by Amaro \textit{et al.} [P. Amaro \textit{et al.}, PRA \textbf{92}, 022514 
(2015)]. We apply this method to estimate the frequency shift of the 1S-3S transition in 
hydrogen and deuterium. Taking into account our experimental conditions, we find a 
frequency shift of less than 1 kHz, that is smaller than our current statistical 
uncertainty.
\end{abstract}


\maketitle

\section{Introduction}

High-resolution spectroscopy plays an important role in testing fundamental theories.
Recently, the proton radius puzzle \cite{puzzle,Carlson} has stimulated a search for overlooked 
systematic effects that could shift atomic transition frequencies. Among such effects is 
the so-called cross-damping effect, or quantum interference. This effect can occur when 
an optically induced atomic transition is detected via the ensuing fluorescence 
\cite{Hessels}. It stems from the presence of neighboring, off-resonant states than can 
be coherently excited along with the resonant transition, and whose decay is detected in 
a non-selective manner. The interference between the different paths leads to a distorted 
and shifted line shape.
This shift of the transition frequency can be important if the off-resonant transitions 
are close enough \cite{Brown}.

Frequency shifts due to quantum interference have been estimated precisely for several 
transitions in muonic hydrogen, deuterium and helium by P. Amaro \textit{et al.} 
\cite{Amaro}, and they have been found to be negligible. However, it is also necessary to 
evaluate these shifts in the case of electronic hydrogen, especially for the 2S-4P 
\cite{Beyer} and 1S-3S transitions.

The two-photon 1S-3S transition of electronic hydrogen is currently studied by the group 
of T.W. Hänsch in Garching \cite{Yost1S3S} and our group in Paris \cite{Galtier}. 
In both experiments, the transition is detected through the Balmer-$\alpha$ fluorescence 
at 656 nm (3S-2P). The cross-damping effect is caused by the presence of the 3D levels, a 
few GHz away from the 3S level, that can be off-resonantly excited and will also decay to 
the 2P levels while emitting photons at 656 nm.
In Garching, the hydrogen atoms are excited by a picosecond pulsed laser. 
Evaluating the quantum interference shift for their measurements \cite{YostQI} required 
the use of the density matrix formalism, leading to complex calculations with many 
coupled equations.
In our experiment, on the contrary, the excitation laser at 205 nm is a continuous-wave 
laser. 
This allows us to use a simpler method, similar to the one developed by P. Amaro 
\textit{et al.} \cite{Amaro}, to estimate the magnitude of the cross-damping effect.

Furthermore, it is also possible to perform the spectroscopy of the 1S-3S transition of 
deuterium using the same experimental setup. In this article, we shall study the quantum 
interference shifts both in hydrogen and in deuterium. 

\section{Theory and calculus}

\subsection{Method}

In order to evaluate the shift due to this quantum interference effect, we follow the 
method described in \cite{Amaro}, adapting it for a two-photon transition and our 
experimental geometry. In the same manner, we can consider the spectroscopy as a two-step 
process equivalent to Raman Stokes scattering, albeit with a two-photon excitation. 

\begin{figure}[h]
\includegraphics{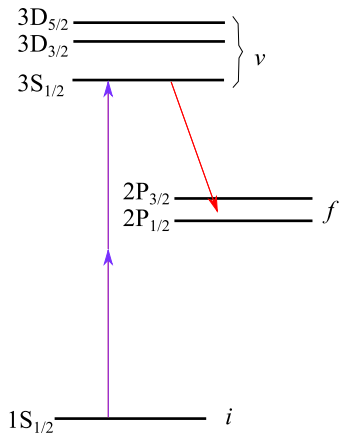}
\caption{The relevant fine-structure energy levels of hydrogen (I=1/2) or deuterium 
(I=1). \label{1S3S}}
\end{figure}

As detailed in Fig. \ref{1S3S}, we will denote $i$ the initial energy level (1S), $\nu$ 
the intermediate level (3S or 3D) of natural linewidth $\Gamma_{\nu}$, and $f$ the final 
level (2P). Table \ref{energies} gives the energies of the relevant hyperfine sublevels.

\begin{table}
\begin{tabular}{ccdcd}
\hline
\hline
&\multicolumn{2}{c}{Hydrogen}&\multicolumn{2}{c}{Deuterium}\\
Level&$F$&\multicolumn{1}{c}{Freq.(MHz)}&$F$&\multicolumn{1}{c}{Freq.(MHz)}\\
\hline\noalign{\smallskip}
3S$_{1/2}$&0&-39.457&1/2&-8.084\\
&1&13.152&3/2&4.042\\
\noalign{\smallskip}
3D$_{3/2}$&1&2927.249&1/2&2929.542\\
&2&2931.458&3/2&2930.027\\
&&&5/2&2930.835\\
\noalign{\smallskip}
3D$_{5/2}$&2&4011.639&3/2&4013.498\\
&3&4014.344&5/2&4013.844\\
&&&7/2&4014.329\\
\hline
\hline
\end{tabular}
\caption{Energies of the $n=3$ hyperfine sublevels relative to the fine structure 
3S$_{1/2}$ level, for hydrogen and deuterium. These energy levels were calculated using 
the NIST database \cite{NIST} and hyperfine structure data published in 
\cite{Karshenboim}. 
The linewidth $\Gamma_{\nu}/2\pi$ is 1.0 MHz for the 3S level and 10.3 MHz for the 3D 
levels. \label{energies}}
\end{table}

Assuming a near-resonant excitation, this scattering process can be described by an 
equation of the Kramers-Heisenberg type, similar to eq. (2) of \cite{Amaro}, in which the 
excitation operator has been replaced by a two-photon operator :
\begin{equation}\label{eq1}
\frac{d\sigma}{d\Omega}\propto\underset{f}{\sum}\left\lvert\underset{\nu}{\sum}~ 
\dfrac{Q_{\nu i}(D_{f\nu})^{*}}{\omega_{\nu i}-2\omega - 
i\Gamma_{\nu}/2}\right\lvert^{2}.
\end{equation}
In this equation, $d\sigma/d\Omega$ is the differential cross section of the scattering 
amplitude, $\omega_{\nu i}$ the transition angular frequency, $\omega$ the laser angular 
frequency, $Q_{\nu i}$ the matrix element of the two-photon excitation operator, and 
$D_{f\nu}$ the dipole matrix element corresponding to the one-photon decay.

The cross-damping effect involves transitions from the same initial state ($J_i=1/2$, 
$F_i$). For a given $F_i$, the sum over $\nu$ can be restricted to the 3S and 3D 
sublevels allowed by the selection rules \cite{Grynberg}:\\
- for the 3S$_{1/2}$ level: $F_{\nu} = F_i$, due to the selection rule $\Delta F = 0$ for 
two-photon transitions between $J=1/2$ states;\\
- for the 3D levels: $\Delta F \leq 2$, with $F_i = 0 \to F_{\nu} = 1$ and $F_i = 1/2 \to 
F_{\nu} = 1/2$ forbidden.

In the present article, we estimate the cross-damping shift for all possible 1S-3S 
hyperfine transitions ($F_i=0$ and 1 for hydrogen, $F_i=1/2$ and 3/2 for deuterium). In 
our current hydrogen experiment, we study the $F_i=1$ transition because the 
1S$_{1/2}^{F=1}$ sublevel is more populated. 

\subsection{Our experimental situation}

We define here the geometry of the scattering process in accordance with 
our experimental situation.
The excitation CW laser at 205 nm is resonant in a Fabry-Perot cavity whose axis is 
horizontal and collinear with the atomic beam. The laser polarization is vertical. 
The 3S-2P fluorescence at 656 nm is collected by an imaging system situated directly 
above the excitation region, and detected by a photo-multiplier. We do not detect the 
polarization of this fluorescence.

\begin{figure}[h]
\includegraphics{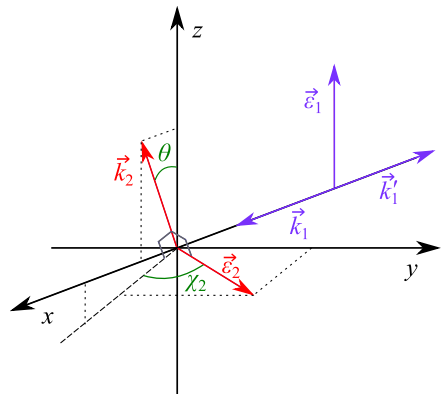}
\caption{The incident photons have opposite wave-vectors $\textbf{k}_1=-\textbf{k}'_1$ 
and the same polarization $\boldsymbol{\varepsilon}_1$. The direction of the wave-vector 
$\textbf{k}_2$ of the scattered photon defines the angle $\theta$. This photon's 
polarization $\boldsymbol{\varepsilon}_2$, which lies in a plane perpendicular to 
$\textbf{k}_2$, makes an angle $\chi_2$ with the scattering plane $xOz$.
\label{polar}}
\end{figure}

Figure \ref{polar} shows the relevant vectors and angles. The two incident photons have 
the same polarization $\boldsymbol{\varepsilon}_1$ (parallel to the $z$ axis), and 
opposite wave-vectors $\textbf{k}_1=-\textbf{k}'_1$ along the $x$ axis. The wave-vector 
$\textbf{k}_2$ of the scattered photon makes an angle $\theta$ with the vertical $z$ 
axis, which is chosen as the quantization axis. As mentioned in \cite{Brown} and 
\cite{Amaro}, the quantum interference effect depends only on this angle $\theta$ between 
the incident polarization and the scattering direction. Without any loss of generality, 
we will assume that this wave-vector $\textbf{k}_2$ is in the plane $xOz$. We also define 
$\chi_2$ as the angle between the scattered photon's polarization 
$\boldsymbol{\varepsilon}_2$ and the plane $xOz$.

The following calculation is done first in the case of a point-like detector situated at 
an angle $\theta$ from the $z$ axis. In order to simulate more closely our experimental 
situation, we will then evaluate the effect for a finite angular aperture of the 
detection system.

\subsection{Details of the calculation}
The polarization vectors of the incident ($\boldsymbol{\varepsilon}_1$) and scattered 
($\boldsymbol{\varepsilon}_2$) photons, as defined above, can be written in a spherical 
basis:
\begin{align}
\varepsilon_{1}^{(\pm 1)}&=0,~~ \varepsilon_{1}^{(0)}=1,\nonumber\\
\varepsilon_{2}^{(\pm 1)}&=\mp\dfrac{(\cos\chi_{2}\cos\theta\pm i \sin\chi_{2})}
{\sqrt{2}},~~ \varepsilon_{2}^{(0)}=-\cos\chi_{2}\sin\theta.
\end{align}

The dipole matrix element is defined as $D_{f\nu}=\boldsymbol{\varepsilon}_{2}.
\textbf{D}_{f\nu}=\langle f|\boldsymbol{\varepsilon}_{2}.\textbf{r}|\nu\rangle$.
We can expand the scalar product in the spherical basis, while taking into account the 
hyperfine structure:
\begin{eqnarray}
D_{F_f m_f J_f}^{F_\nu m_\nu J_\nu}=\underset{\lambda=-1}{\overset{1}
{\sum}}&&(-1)^{\lambda}\varepsilon_{2}^{(-\lambda)}\nonumber\\&&\langle n_f L_f F_f m_f 
J_f  | r_{\lambda} | n_{\nu}L_{\nu} F_\nu m_\nu J_\nu \rangle\label{D}.
\end{eqnarray}

The two-photon matrix element is expressed as:
\begin{eqnarray}
Q_{\nu i}&=&\left| \underset{r}{\sum}\frac{(\boldsymbol{\varepsilon}_{1}.\textbf{D}_{\nu 
r})(\boldsymbol{\varepsilon}_{1}.\textbf{D}_{ri})}{\omega-\omega_{ri}}\right|
^{2}\nonumber\\
&=&\left| \underset{r}{\sum}\frac{\langle\nu|\boldsymbol{\varepsilon}_{1}.\textbf{r}|
r\rangle\langle r|\boldsymbol{\varepsilon}_{1}.\textbf{r}|i\rangle}
{\omega-\omega_{ri}}\right|^{2}.
\end{eqnarray}
It can also be written as the matrix element of a $k^{\text{th}}$-order tensor operator 
$\textsf{\textbf{T}}^{(k)}$, with $k=0$ for 1S-3S, $k=2$ for 1S-3D \cite{Grynberg}. Since 
the incident polarization $\boldsymbol{\varepsilon}_{1}$ is along the quantization axis 
(this implies $m_\nu=m_i$), we simply have:
\begin{equation}
Q_{F_\nu m_\nu J_\nu}^{F_i m_i J_i}=\langle  n_\nu L_\nu F_\nu m_\nu J_\nu | T_{0}^{(k)} 
| n_i L_i  F_i m_i J_i \rangle\label{Q}.
\end{equation}

Defining $\textsf{\textbf{T}}^{(1)} = \textbf{r}$, the matrix elements on the right-hand 
side of eqs. \eqref{D} and \eqref{Q} can be simplified by introducing the reduced matrix 
element, then successively decoupling the angular momenta to separate radial and angular 
parts, using the following usual relations \cite{Edmonds}: 
\begin{widetext}
\begin{align}
&\langle n' L' F' m' J'  | T_{\lambda}^{(k)} | nL F mJ \rangle=(-1)^{F'-m'}\left( 
\begin{array}{ccc}
F' & k & F \\
-m' & \lambda & m \end{array}\right) \langle n' L' F'  J' || \textsf{\textbf{T}}^{(k)} || 
nL F J \rangle \nonumber,\\
&\langle n' L' F'  J'  || \textsf{\textbf{T}}^{(k)} || nL F J \rangle 
=(-1)^{J'+I+F+k}\sqrt{[F,F']}\left\lbrace\begin{array}{ccc}
J' & F' & I \\
F & J & k \end{array}\right\rbrace \langle n' L' J' || \textsf{\textbf{T}}^{(k)} || nL J 
\rangle \nonumber,\\
&\langle n' L' J'  || \textsf{\textbf{T}}^{(k)} || nL J 
\rangle=(-1)^{L'+S+J+k}\sqrt{[J,J']}\left\lbrace\begin{array}{ccc}
L' & J' & S \\
J & L & k \end{array}\right\rbrace\langle n' L' || \textsf{\textbf{T}}^{(k)} || nL 
\rangle,
\end{align}
\end{widetext}
with the notation $[J,J']=(2J+1)(2J'+1)$. One obtains
\begin{align}
D_{F_f m_f J_f}^{F_\nu m_\nu J_\nu}
&=\langle n_{f}L_{f} || \textbf{r} || n_{\nu}L_{\nu} \rangle \times \underset{\lambda=-1}
{\overset{1}{\sum}}(-1)^{\lambda}\varepsilon_{2}^{(-\lambda)} A^{f 
\nu}_\lambda(1)\nonumber,\\
Q_{F_\nu m_\nu J_\nu}^{F_i m_i J_i}
&=\langle n_\nu L_\nu || \textsf{\textbf{T}}^{(k)} || n_i L_i \rangle \times  A^{\nu 
i}_0(k),
\end{align}
where we have introduced the angular coefficient 
$A_{\lambda}(k)$ for a $k^{\text{th}}$-order tensor operator $\textsf{\textbf{T}}^{(k)}$:
\begin{align}
A_{\lambda}(k)&=(-1)^{F'-m'}\left( \begin{array}{ccc}
F' & k & F \\
-m' & \lambda & m \end{array}\right) \nonumber\\
\times &(-1)^{J'+I+F+k}\sqrt{[F,F']}\left\lbrace\begin{array}{ccc}
J' & F' & I \\
F & J & k \end{array}\right\rbrace \nonumber\\
\times &(-1)^{L'+S+J+k}\sqrt{[J,J']}\left\lbrace\begin{array}{ccc}
L' & J' & S \\
J & L & k \end{array}\right\rbrace.
\end{align}
It should be noted that $A_0(0)=1$, as there is no angular coefficient for the 1S-3S 
excitation.

One can then rearrange the terms to separate radial and angular parts: 
\begin{equation}
Q_{F_\nu m_\nu J_\nu}^{F_i m_i J_i}\left(D_{F_f m_f J_f}^{F_\nu m_\nu 
J_\nu}\right)^*=S_{f\nu i}~\Omega_{J_{i}J_{\nu}J_{f}}^{F_{i}F_{\nu}F_{f}},
\end{equation}
with
\begin{eqnarray}
S_{f\nu i}=\langle n_{f}L_{f} || \textbf{r} || n_{\nu}L_{\nu} \rangle\langle 
n_{\nu}L_{\nu} || \textsf{\textbf{T}}^{(k)} || n_{i}L_{i} \rangle, \label{radial}\\
\Omega_{J_{i}J_{\nu}J_{f}}^{F_{i}F_{\nu}F_{f}}=\underset{m_\nu,\lambda}{\sum}
(-1)^{\lambda}\left(\varepsilon_{2}^{(-\lambda)}\right)^{*}
A_0(k)A_{\lambda}(1).
\end{eqnarray}

Replacing in eq. (\ref{eq1}), one obtains:
\begin{equation}
\frac{d\sigma}{d\Omega}\propto\underset{\begin{subarray}{c}
F_{f},J_{f}, \\ m_{i},m_{f},\boldsymbol{\varepsilon}_2 \end{subarray}}{\sum} 
~\left\lvert\underset{F_\nu,J_\nu}{\sum}~ \dfrac{S_{f\nu 
i}~\Omega_{J_{i}J_{\nu}J_{f}}^{F_{i}F_{\nu}F_{f}}}{\omega_{\nu i}-2\omega - 
i\Gamma_{\nu}/2}\right\lvert^{2}.
\end{equation}
It is necessary to sum over $\boldsymbol{\varepsilon}_2$ because the polarization of the 
scattered photon is not detected.

As in \cite{Amaro}, the terms can be further rearranged to show direct and cross terms:
\begin{widetext}
\begin{equation}
\label{signal}
\frac{d\sigma}{d\Omega}\propto\underset{F_{\nu},J_{\nu}}{\sum}\dfrac{ S_{f\nu 
i}^{2}~\Lambda_{J_{i}J_{\nu}}^{F_{i}F_{\nu}}}{(\omega_{\nu 
i}-2\omega)^{2}+(\Gamma_{\nu}/2)^{2}}~+~\text{Re}\left[ 
\underset{(F'_{\nu},J'_{\nu})>(F_{\nu},J_{\nu})}{\sum} \dfrac{S_{f\nu i}~S_{f\nu' 
i}~\Xi_{J_{i}J_{\nu}J_{\nu'}}^{F_{i}F_{\nu}F_{\nu'}}}{(\omega_{\nu i}-2\omega-
i\Gamma_{\nu}/2)(\omega_{\nu' i}-2\omega+i\Gamma_{\nu'}/2)}\right]
\end{equation}
where we have defined
\begin{equation}
\Lambda_{J_{i}J_{\nu}}^{F_{i}F_{\nu}}= \underset{\begin{subarray}{c}
F_{f},J_{f}, \\ m_{i},m_{f},\boldsymbol{\varepsilon}_2 \end{subarray}}{\sum} \left| 
\Omega_{J_{i}J_{\nu}J_{f}}^{F_{i}F_{\nu}F_{f}} \right|^{2}~~~\text{and}~~~ 
\Xi_{J_{i}J_{\nu}J_{\nu'}}^{F_{i}F_{\nu}F_{\nu'}} = 2 \text{Re}\left[ 
\underset{\begin{subarray}{c}
F_{f},J_{f}, \\ m_{i},m_{f},\boldsymbol{\varepsilon}_2 \end{subarray}}{\sum}  
\Omega_{J_{i}J_{\nu}J_{f}}^{F_{i}F_{\nu}F_{f}} 
\left(\Omega_{J_{i}J_{\nu'}J_{f}}^{F_{i}F_{\nu'}F_{f}}\right)^{*}\right].
\end{equation}
\end{widetext}

\subsection{Radial part}

The two matrix elements in eq. \eqref{radial} can be evaluated in the following way.

$\langle n_{f}L_{f} || \textbf{r} || n_{\nu}L_{\nu}\rangle$ is the well-known reduced 
matrix element of the radial operator \textbf{r} and can be easily calculated using the 
Wigner-Eckart theorem: 
\begin{eqnarray}
\langle nLm|& r_{\lambda} | &n'L'm'\rangle\nonumber\\
&=&(-1)^{L-m}\left( \begin{array}{ccc}
L & 1 & L' \\
-m & \lambda & m \end{array}\right)\langle nL || \textbf{r} || n'L'\rangle.
\end{eqnarray}
For example, defining $\psi_{nLm}$ as the usual electronic wave function of hydrogen, one 
has:
\begin{eqnarray}
\langle 2\text{P}0| z | 3\text{D}0\rangle &=&(-1)^{1-0}\left( \begin{array}{ccc}
1 & 1 & 2 \\
0 & 0 & 0 \end{array}\right)\langle 2\text{P} || \textbf{r} || 
3\text{D}\rangle\nonumber\\
&=&\int\psi_{210}^{*}(r)~z~
\psi_{320}(r)~d^{3}r,
\end{eqnarray}
where the integral is calculated over the whole space.

The two-photon matrix element has been calculated by M. Haas \textit{et al.} \cite{Haas}. It 
is given by:
\begin{equation}
\langle n_{\nu}L_{\nu} || \textsf{\textbf{T}}^{(k)} || n_{i}L_{i} 
\rangle=-\frac{2hc\epsilon_0}{e^{2}}\times\beta_{ge}^{(k)},
\end{equation} 
where the coefficients $\beta_{ge}^{(k)}=\beta_{ge}$ for 1S-3S and $\beta_{ge}^{(2)}$ for 
1S-3D are given in tables II and III of \cite{Haas}. These coefficients are given in 
Hz(W/m$^{2}$)$^{-1}$. 

In our case, the radial part is  
\begin{align}
&\langle 3\text{S} || \textsf{\textbf{T}}^{(0)} || 1\text{S} \rangle=1.00333\times C,
\nonumber\\ 
&\langle 3\text{D} || \textsf{\textbf{T}}^{(2)} || 1\text{S} \rangle=-6.16579\times C,
\nonumber\\ &\langle 2\text{P} || \textbf{r} || 3\text{S} \rangle=0.938404\times a_0,
\nonumber\\ &\langle 2\text{P} || \textbf{r} || 3\text{D} \rangle=-6.71467\times a_0,
\end{align} 
where $C=-10^{-5}\times \frac{2hc\epsilon_0}{e^{2}}$ and $a_0$ is the Bohr radius. Both 
constants are global factors and we do not take them into account.

In the numerical calculations, we then simply used:
\begin{equation}
S_{f\nu i}=\left\lbrace\begin{array}{cc} 1.00333\times0.938404&\text{for 
}\nu=\text{3S}\\-6.16579\times(-6.71467)&\text{for }\nu=\text{3D} \end{array}\right. 
\label{radialnum}
\end{equation}

\subsection{Angular part}

As noted earlier, the quantum interference effect depends only on the angle $\theta$ 
between the incident polarization and the scattering direction. Hence, the coefficients 
$\Lambda$ and $\Xi$ have a simple angular dependence and can be parametrized as follows:
\begin{eqnarray}
&&\Lambda_{J_{i}J_{\nu}}^{F_{i}F_{\nu}}(\theta)=a_{0}+a_{2}P_{2}(\cos \theta),\nonumber\\
&&\Xi_{J_{i}J_{\nu}J_{\nu'}}^{F_{i}F_{\nu}F_{\nu'}}(\theta)=b_{2}P_{2}(\cos \theta),
\label{param2}
\end{eqnarray}
where $P_{2}$ is the second-order Legendre polynomial: $P_{2}(x)=(3x^{2}-1)/2$.

Table \ref{coeff1S3S} gives the coefficients of this parametrization for hydrogen: direct 
terms for each hyperfine transition, and cross terms between the 1S-3S transition and the 
1S-3D transitions. 
In deuterium, the hyperfine structure is different but the method developed above can be 
directly applied: the radial part is the same (eq. \eqref{radialnum}), and the angular 
part should be changed accordingly (Table \ref{coeff1S3Sdeut}). 

The cross terms between 3D levels play a negligible role in the distortion and shifting 
of the 1S-3S line. They are not included in these tables but are given in the Appendix.

\begin{table}[h]
\begin{center}
\begin{tabular}{ccccccc}
\hline 
\hline
$F_i$ & $L_{\nu}$ & $F_{\nu}$ & $J_{\nu}$ & $a_{0}$ & $a_{2}$  & $b_{2}$\\ 
\hline 
0 & 0 & 1 & 1/2 & 2/3 & 0 & \\ 
 & 2 & 2 & 3/2 & 4/375 & -7/1875 & $4\sqrt{2}/75$\\ 
 & 2 & 2 & 5/2 & 2/125 & -4/625 & $2\sqrt{2}/25$\\ 
\noalign{\smallskip}
1 & 0 & 1 & 1/2 & 2 & 0 & \\ 
 & 2 & 1 & 3/2 & 2/125 & -7/2500  & $2\sqrt{2}/25$\\ 
 & 2 & 2 & 3/2 & 2/125 & -7/2500 & $2\sqrt{2}/25$\\ 
 & 2 & 2 & 5/2 & 4/375 & -4/1875 & $4\sqrt{2}/75$\\ 
 & 2 & 3 & 5/2 & 14/375 & -8/625 & $14\sqrt{2}/75$\\ 
\hline
\hline
\end{tabular} 
\caption{Angular coefficients for hydrogen, $F_i=0$ and 1. \label{coeff1S3S}}
\end{center}
\end{table}

\begin{table}[h]
\begin{center}
\begin{tabular}{ccccccc}
\hline 
\hline
$F_i$ & $L_{\nu}$ & $F_{\nu}$ & $J_{\nu}$ & $a_{0}$ & $a_{2}$  & $b_{2}$\\ 
\hline 
1/2 & 0 & 1/2 & 1/2 & 4/3 & 0 & \\ 
 & 2 & 3/2 & 3/2 & 8/1875 & -14/46875 & $8\sqrt{2}/375$\\ 
 & 2 & 5/2 & 3/2 & 32/1875 & -224/46875 & $32\sqrt{2}/375$\\ 
 & 2 & 3/2 & 5/2 & 32/1875 & -224/46875 & $32\sqrt{2}/375$\\ 
 & 2 & 5/2 & 5/2 & 28/1875 & -184/46875 & $28\sqrt{2}/375$\\ 
\noalign{\smallskip}
3/2 & 0 & 3/2 & 1/2 & 8/3 & 0 & \\ 
 & 2 & 1/2 & 3/2 & 4/375 & 0 & $4\sqrt{2}/75$\\ 
 & 2 & 3/2 & 3/2 & 32/1875 & 0 & $32\sqrt{2}/375$\\ 
 & 2 & 5/2 & 3/2 & 28/1875 & -14/9375 & $28\sqrt{2}/375$\\ 
 & 2 & 3/2 & 5/2 & 8/1875 & 0 & $8\sqrt{2}/375$\\ 
 & 2 & 5/2 & 5/2 & 32/1875 & -436/459375 & $32\sqrt{2}/375$\\ 
 & 2 & 7/2 & 5/2 & 16/375 & -16/1225 & $16\sqrt{2}/75$\\ 
\hline
\hline
\end{tabular} 
\caption{Angular coefficients for deuterium, $F_i=1/2$ and 3/2. \label{coeff1S3Sdeut}}
\end{center}
\end{table}

\section{Results}

In order to estimate the frequency shift due to the cross-damping effect, we calculate a 
simulated signal taking into account the direct and cross terms using eq. \eqref{signal}. 
We then fit the 1S-3S line with a simple Lorentzian function, leaving all fit parameters 
(position, width, amplitude) free. 
The shift is defined here as the difference between the position given by the fit and the 
theoretical position used in the calculation.

We do not add any noise to the simulated spectrum; in our experiment, there is a rather 
large background so the noise can be approximated by a white noise. We have checked that 
adding a white noise to the simulated signal does not significantly change the result of 
the fit.

\subsection{Point-like detector}
\begin{figure}
\includegraphics{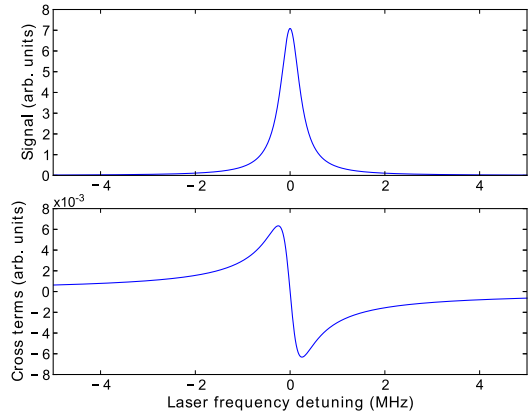}
\caption[caption]{(a) Simulated 1S-3S signal for hydrogen, $F_i=1$, $\theta=0$. (b) Sum 
of the cross terms;
the arbitrary units are the same as in (a), but the vertical scale is amplified by a 
factor of 500.\label{sig}}
\end{figure}

Figure \ref{sig}(a) shows the simulated signal for hydrogen, $F_i=1$, in the case of a 
point-like detector situated directly above the excitation point ($\theta=0$). The second 
term on the right-hand side of eq. \eqref{signal} is the signature of quantum 
interference, and is represented in Fig. \ref{sig}(b). Its dispersion shape is 
responsible for the shift of the transition frequency. 
All the results given below are shifts of the laser frequency $\omega/2\pi$, and differ 
from the atomic transition frequency shifts by a factor of two.
\begin{figure}
\includegraphics{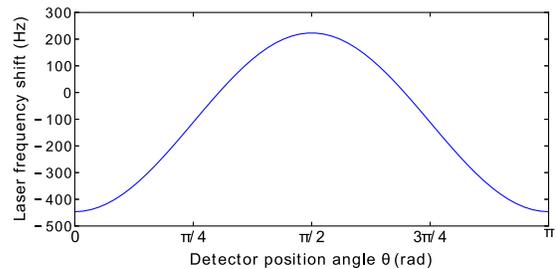}
\caption{Shift of the laser frequency as a function of detector position angle $\theta$.
\label{fig}}
\end{figure}

Figure \ref{fig} shows the frequency shift as a function of the position of a point-like 
detector. The shift is maximal for $\theta=0$, and is proportional to $P_{2}(\cos 
\theta)$, having the same angular dependence as the amplitude of the cross terms. This 
fact is not surprising, since the shift is very small compared to the natural linewidth, 
and can be expected to vary linearly with the amplitude of the cross terms.

This figure is comparable to the results of D. Yost \textit{et al.} (Fig. 5 of 
\cite{YostQI}), that were calculated using a completely different method in which the 
continuous excitation was treated as a special case.

\begin{table}
\begin{tabular}{ccc}
\hline
\hline
&$F_i$&Shift (Hz)\\
\hline
H & 0 & $-$440 \\
  & 1 & $-$446 \\ 
\noalign{\smallskip}
D & 1/2 & $-$444 \\ 
  & 3/2 & $-$445 \\ 
\hline
\hline
\end{tabular}
\caption{Calculated shift for $\theta=0$, in Hz. \label{shifts}}
\end{table}

Table \ref{shifts} gives the maximal shift, calculated for $\theta=0$, for the four 
possible hyperfine transitions. It is interesting to notice that we find very similar 
shifts for the different cases. This is due to the fact that the hyperfine structure of 
the 3D levels is not resolved because it is smaller than the natural linewidth of these 
levels. The frequency shift is thus at most of $-$0.45 kHz for all 1S-3S transitions; we 
also find this result if we ignore the hyperfine structure in the calculations.

One can also compare this shift to a naive estimate derived from the simplified case of a 
three-level atom. The calculation of the first term in eq. (25) of \cite{Hessels} would 
give, with $\Gamma = 1$ MHz and $\Delta \approx 3000$ MHz: 
\begin{equation}
\Gamma^{2}/4\Delta\approx0.08~\rm kHz.
\end{equation}
In fact, this gives the atomic frequency shift due to a single cross term between excited 
levels of linewidth $\Gamma$ and separated by $\Delta$, assuming that the cross term and 
direct term have the same amplitude. 
In our case, there are several cross-terms, and as we can see in eq. \eqref{signal}, 
these cross terms all have different amplitudes; thus, we should take into account the 
amplitude ratio between each cross  term and the direct term, and sum over all 3D sublevels $\nu'$ interfering with the 3S level $\nu$, in order to calculate the total atomic 
frequency shift:
\begin{equation}
\delta(2\omega)\approx\underset{\nu'}{\sum}\left[\frac{\Gamma_{\nu}^{2}}{4(\omega_{\nu 
i}-\omega_{\nu'i})}\times\frac{S_{f\nu' 
i}\times\Xi_{J_{i}J_{\nu}J_{\nu'}}^{F_{i}F_{\nu}F_{\nu'}}(\theta)}{S_{f\nu 
i}\times\Lambda_{J_{i}J_{\nu}}^{F_{i}F_{\nu}}}\right].
\end{equation}
For $\theta=0$, this equation gives $\delta(2\omega)\approx-0.45$ kHz, which is indeed a 
very good estimate of the shift.

\subsection{Extended detector}

\begin{figure}
\includegraphics{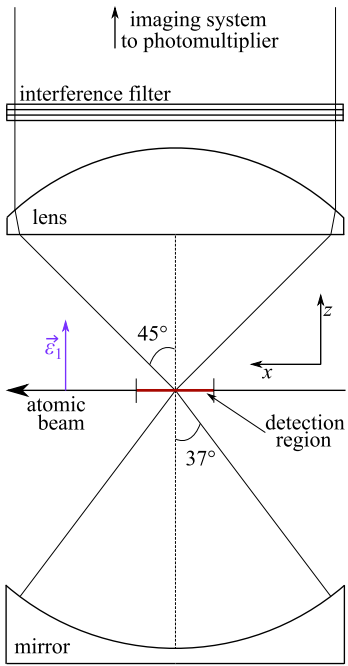}
\caption{Side view of the fluorescence collection system.\label{imag}}
\end{figure}

In order to simulate more closely our experiment, we can integrate the signal over the 
angular aperture of our imaging system. The point-like detector case for $\theta=0$ gives 
an upper bound for the frequency shift; any integration over this angle will only reduce 
the effect. Furthermore, integrating over the whole space cancels the effect altogether.

The fluorescence collection system is shown in Fig. \ref{imag}. The scattered photons are 
collected through an aspheric lens of radius 25 mm and an interference filter at 656 nm. 
A spherical metallic mirror, having the same radius as the lens and situated below the 
excitation region, increases the solid angle of detection by redirecting photons emitted 
downwards. The 10$^{\circ}$ acceptance angle of the interference filter limits the length 
of the detection region along the atomic beam, which is then a segment of length 12 mm 
centered on the waist of the 205 nm Fabry-Perot cavity. The center of this detection 
region is the focal point of the lens as well as the center of curvature of the spherical 
mirror. 

Let us assume for now that the detection region is infinitesimal and centered: in this 
situation, only photons emitted at the center of the cavity are detected. We can first 
integrate the simulated signal over the upper part of the collection system:
\begin{equation}
Signal=\int_{0}^{\theta_{max}}f(\theta)\times 2\pi \sin(\theta)d\theta, \label{integrate}
\end{equation}
where $f(\theta)$ is the right-hand side of eq. \eqref{signal}, and $\theta_{max}$ is the 
half angle of the detection cone.
With $\theta_{max}=45^\circ$ defined by the diameter of the lens, equation 
\eqref{integrate} leads to a laser frequency shift of $-$0.27 kHz.

Then, it is possible to calculate the signal for a given position of the emission point 
along the detection region. The angular acceptance of the filter can be approximated by a 
step function of the incident angle, so that the distribution of the emission points is 
assumed to be uniform along the segment. Integrating over the length of the detection 
region does not change the result significantly ($<$1 Hz). 

We can thus simply add to the previous signal of eq. \eqref{integrate} the integral over 
the downwards-emitted photons reflected by the spherical mirror, with an opening half-
angle of 37$^{\circ}$, neglecting the losses due to the reflection on the mirror:
\begin{eqnarray}
Signal=\int_{0}^{45^\circ}&f(\theta)&\times 2\pi \sin(\theta)d\theta\nonumber\\&+& 
\int_{0}^{37^\circ}f(\theta)\times 2\pi \sin(\theta)d\theta \label{integrate2}.
\end{eqnarray}

This results in a frequency shift of $-$0.29 kHz.

\section{Conclusion}

In this article, we have estimated the frequency shift, due to the cross-damping effect, 
of the 1S-3S transition of hydrogen and deuterium, in the conditions of the experiment in 
progress in our group. 
This shift is similar for both isotopes, and depends on the angle at which the 
fluorescence photon is emitted with respect to the polarization of the incident laser 
light. The maximal shift of the laser frequency is of $-$0.45 kHz, assuming a point-like 
detector situated at the vertical of the excitation point.
After taking into account the actual geometry of our detection system, we found a laser 
frequency shift of $-$0.29 kHz. This corresponds to a shift of the atomic frequency of 
$-$0.58 kHz, that is smaller than the current statistical uncertainty (2.2 kHz 
\cite{Galtier}) of our measurements.

\begin{acknowledgments}
This work is supported by the cluster of excellence FIRST-TF via a public grant from the 
French National Research Agency (ANR) as part of the ‘Investissements d’Avenir’ program 
(reference: ANR-10-LABX-48). J.-Ph. K. acknowledges support as a fellow of Institut 
Universitaire de France.
\end{acknowledgments} 

\appendix*
\section{}

Tables \ref{coeff1S3D} and \ref{coeff1S3DD} present the coefficient $b_{2}$, as defined 
in eq. \eqref{param2}, for the cross terms between the different 3D hyperfine sublevels. 
These cross terms do not play any significant role in shifting the 1S-3S line.
\begin{table}[h]
\begin{tabular}{cccccc}
\hline
\hline 
$F_i$ & $F_{\nu}$ & $J_{\nu}$ & $F_{\nu'}$ & $J_{\nu'}$ & $b_{2}$\\ 
\hline 
0 & 2 & 3/2 & 2 & 5/2 & -2/625\\ 
\noalign{\smallskip}
1 & 1 & 3/2 & 2 & 3/2 & -7/1250\\ 
 & 1 & 3/2 & 2 & 5/2 & -7/1875\\ 
 & 1 & 3/2 & 3 & 5/2 & -2/1875\\ 
 & 2 & 3/2 & 2 & 5/2 & 1/625\\ 
 & 2 & 3/2 & 3 & 5/2 & -4/625\\ 
 & 2 & 5/2 & 3 & 5/2 & -8/1875\\ 
\hline
\hline 
\end{tabular} 
\caption{Angular coefficients of cross terms between 3D sublevels of hydrogen.
\label{coeff1S3D}}
\end{table}

\begin{table}
\begin{tabular}{cccccc}
\hline 
\hline 
$F_i$ & $F_{\nu}$ & $J_{\nu}$ & $F_{\nu'}$ & $J_{\nu'}$ & $b_{2}$\\ 
\hline 
1/2 & 3/2 & 3/2 & 5/2 & 3/2 & -112/46875\\ 
 & 3/2 & 3/2 & 3/2 & 5/2 & -112/46875\\
 & 3/2 & 3/2 & 5/2 & 5/2 & 52/46875\\
 & 5/2 & 3/2 & 3/2 & 5/2 & -16/15625\\
 & 5/2 & 3/2 & 5/2 & 5/2 & -64/15625\\
 & 3/2 & 5/2 & 5/2 & 5/2 & -64/15625\\ 
\noalign{\smallskip}
3/2 & 1/2 & 3/2 & 3/2 & 3/2 & -56/9375\\ 
 & 1/2 & 3/2 & 5/2 & 3/2 & -14/9375\\
 & 1/2 & 3/2 & 3/2 & 5/2 & -14/9375\\
 & 1/2 & 3/2 & 5/2 & 5/2 & -16/9375\\
 & 1/2 & 3/2 & 7/2 & 5/2 & 0\\
 & 3/2 & 3/2 & 5/2 & 3/2 & -56/9375\\
 & 3/2 & 3/2 & 3/2 & 5/2 & 0\\
 & 3/2 & 3/2 & 5/2 & 5/2 & -208/65625\\
 & 3/2 & 3/2 & 7/2 & 5/2 & -128/65625\\
 & 5/2 & 3/2 & 3/2 & 5/2 & 2/3125\\
 & 5/2 & 3/2 & 5/2 & 5/2 & 32/21875\\
 & 5/2 & 3/2 & 7/2 & 5/2 & -144/21875\\
 & 3/2 & 5/2 & 5/2 & 5/2 & -64/21875\\
 & 3/2 & 5/2 & 7/2 & 5/2 & -32/65625\\
 & 5/2 & 5/2 & 7/2 & 5/2 & -1152/153125\\
\hline
\hline 
\end{tabular} 
\caption{Angular coefficients of cross terms between 3D sublevels of deuterium.
\label{coeff1S3DD}}
\end{table}

\end{document}